\documentstyle[12pt]{article}

\begin{document}
\begin{titlepage}
\title{The Causal Interpretation of Quantum Mechanics and The
Singularity Problem in Quantum Cosmology}
\author{{\bf J. Acacio de Barros}\thanks{e-mail address: 
acacio@fisica.ufjf.br} ${}^{a}$\\and\\
{\bf N. Pinto-Neto}\thanks{e-mail address:
nen@lca1.drp.cbpf.br}${}^{\;\;b}$\\ 
${}^{a}${\it Departamento de F\'{\i}sica - ICE}\\
{\it Universidade Federal de Juiz de Fora}\\
{\it 36036-330, Juiz de Fora, MG, Brazil}\\
${}^{b}${\it Centro Brasileiro de Pesquisas F\'{\i}sicas/Lafex}\\
{\it Rua Xavier Sigaud, 150 - Urca}\\
{\it 22290-180, Rio de Janeiro, RJ,Brazil}}
\maketitle
\newpage
\begin{abstract}
We apply the causal interpretation of quantum mechanics to homogeneous
quantum cosmology and show that the quantum theory is independent of any
time-gauge choice and there is no issue of time. 
We exemplify this result by studying a particular minisuperspace model
where the quantum potential driven by a prescribed quantum state prevents
the formation of the classical singularity, independently on the choice
of the lapse function. This means that the fast-slow-time gauge conjecture
is irrelevant within the framework of the causal interpretation of
quantum cosmology.

\vspace{0.7cm}
PACS number(s): 98.80.H, 03.65.Bz
\end{abstract}
\end{titlepage}

\section{Introduction}

The singularity theorems \cite{he} show that, under reasonable physical
assumptions, the Universe has developed an initial singularity,
and will develop future singularities in the form of black holes
and, perhaps, of a big crunch. Until now, singularities are out of the
scope of any physical theory. If we assume that a physical 
theory can describe the whole Universe at 
every instant, even at its possible moment of creation
(which is the best attitude because it is the only 
way to seek  the limits of physical science), then it is necessary that
the `reasonable physical assumptions'  of the theorems be
not valid under extreme situations of very high energy density and curvature.
We may say that general relativity, and/or any other matter field theory,
must be changed under these extreme conditions. One good point of
view is to think that quantum gravitational effects become
important, eliminating the singularities that should appear classically,
similarly to what happens with the quantum atom.
We should then construct a quantum theory of gravitation,
apply it to cosmology, and see if it works. However, there is 
no established theory of quantum gravity. Furthermore, any quantum
theory when applied to cosmology presents new profound conceptual 
problems. How can we
apply the standard probabilistic Copenhaguen interpretation to a single 
system?  Where in a quantum Universe
can we find a classical domain where we could construct our classical
measuring apparatus to test and give sense to the quantum theory? 
Who are the observers of the whole Universe?
This is not a problem of quantum gravity alone
because there is no problem with the concept of an ensemble
of black holes and a classical domain outside it. 
Finally, in quantum mechanics, time is not treated
as an observable (hermitean operator) but as an external evolution
parameter (c-number). In the quantum cosmology of a closed
universe, there is no place for an external parameter. So, what
happens with time? Which internal variable will give a sense of evolution
of the quantum states? 

In this paper we will close our attention to the interpretation and time
issues in order to study the singularity problem in quantum cosmology. 
The difficult
technical problems coming from the quantization of the full gravitational
field will be circumvented by taking advantage of minisuperspace models
which restrict the gravitational and matter fields to be homogeneous.
In these models, all but a finite number of degrees of freedom are
frozen out alleviating considerably the technical problems. 

In the 
framework of these minisuperspace models, a number of papers have been
written showing how the issue of time is important for the singularity
problem: different choices of time imply different quantum theories,
some of them still presenting singularities, others not \cite{isham,lemos}.
The interpretation adopted is the conventional probabilistic one.
Here, we will adopt a non-probabilistic interpretation 
to quantum cosmology which circumvents the measurement problem because
it is an ontological interpretation of quantum mechanics: it
is not necessary to have a measuring apparatus or a classical domain
in order to recover physical reality; it is there ``ab initio". 
It is the causal interpretation of quantum
mechanics \cite{boh,hol}. We will apply this interpretation to
the minisuperspace models of homogeneous gravitational and matter fields
mentioned above, and show that the question about the persistency of
the singularities at the quantum level does not depend on the choice
of time but only on the quantum state of the system. A particular example
will be exhibited to bring home this fact. 

This paper is organized as follows: in the next section we make a summary
of the causal interpretation. In section 3, we apply this interpretation
to quantum cosmology, and show that, for the minisuperspace models of
homogeneous gravitational and matter fields, the quantum theory is independent
on the choice of time. We also call attention to the fact that this result
may no longer be valid for
inhomogeneous fields. 
In section 4, we present a particular minisuperspace example where the
classical singularities can be removed by a choice of the quantum state,
and show that this result does not depend on the choice of time. We end 
with some comments and conclusions.

\section{The causal interpretation of quantum mechanics}

In this section, we will review the ontological interpretation
of quantum mechanics, and apply it to quantum cosmology. 
Let us begin with the Schr\"{o}dinger equation,
in the coordinate representation, for a non-relativistic particle
with the hamiltonian $H = p^2 / 2m + V(x)$:

\begin{equation}
\label{bsc}
i \hbar \frac{d \Psi (x,t)}{d t} = [-\frac{\hbar ^2}{2m} \nabla ^2 +
V(x)] \Psi (x,t) .
\end{equation}
Writing $\Psi = R \exp (iS/\hbar)$, and substituting it into (\ref{bsc}),
we obtain the following equations:
\begin{equation}
\label{bqp}
\frac{\partial S}{\partial t}+\frac{(\nabla S)^2}{2m} + V
-\frac{\hbar ^2}{2m}\frac{\nabla ^2 R}{R} = 0 ,
\end{equation}
\begin{equation}
\label{bpr}
\frac{\partial R^2}{\partial t}+\nabla .(R^2 \frac{\nabla S}{m}) = 0 .
\end{equation}

The usual probabilistic interpretation takes equation (\ref{bpr}) and
understands it as a continuity equation for the probability density $R^2$
for finding the particle at position $x$ and time $t$.
All physical information about the system is contained 
in $R^2$, and the total phase $S$ of the wave function is 
completely irrelevant. 
In this interpretation, nothing is
said about $S$ and its evolution equation (\ref{bqp}). 
However, examining equation (\ref{bpr}), we can see
that $\nabla S /m$ may be interpreted as a velocity field, suggesting
the identification $p=\nabla S$. Hence, we can look to
equation (\ref{bqp}) as a Hamilton-Jacobi
equation for the particle with the extra potential term
$-\hbar ^2 \nabla ^2 R /2m R$. 

After this preliminary, let us introduce the ontological interpretation
of quantum mechanics, which is based on the {\it two} equations
(\ref{bqp}) and (\ref{bpr}), and not only in the last one as it is
the Copenhaguen interpretation:

i) A quantum system is composed of a particle {\it and} a field $\Psi$
(obeying the Schr\"{o}dinger equation (\ref{bsc})), each one having its own
physical reality.

ii) The quantum particles follow trajectories $x(t)$, {\it independent
on observations}. Hence, in this interpretation, we can talk about
trajectories of quantum particles, contrary to the Copenhaguen interpretation
where only positions at one instant of time have a physical meaning.

iii) The momentum of the particle is $p=\nabla S$.

iv) For a statistical ensemble of particles in the same quantum
field $\Psi$, the probability density is $P=R^2$. Equation (\ref{bpr})
guarantees the conservation of $P$.

Let us make some comments:

a) Equation (\ref{bqp}) can now be interpreted as a Hamilton-Jacobi 
type equation for a particle submited to an external potential which 
is the classical
potential plus a new quantum potential  
\begin{equation}
\label{qp}
Q \equiv -\frac{\hbar ^2}{2m}\frac{\nabla ^2 R}{R} . 
\end{equation}
Hence, the particle trajectory $x(t)$ satisfies the equation of motion
\begin{equation}
\label{beqm}
m \frac{d^2 x}{d t^2} = -\nabla V - \nabla Q .  
\end{equation}

b) Even in the regions where $\Psi$ is very small, the quantum potential
can be very high, as we can see from equation (\ref{qp}). It depends
only on the form of $\Psi$, not on its absolute value. This fact
brings home the non-local and contextual character of the quantum 
potential\footnote{This fact becomes evident when we generalize the causal
interpretation to a many particle system.}. This is
very important because Bell's inequalities together
with Aspect's experiments show that, in general, a
quantum theory must be either non-local or non-ontological. As
Bohm's interpretation is ontological, it must be non-local, as it is.
The quantum potential is responsible for the quantum effects. 

c) This interpretation can be applied to a single particle. In this
case, equation (\ref{bpr}) is just an equation to determine the function $R$,
which forms the quantum potential acting on the particle via equation 
(\ref{beqm}). The function $R^2$ does not need to be 
interpreted as a probability density and hence
needs not be normalized. The interpretation of $R^2$ as a probability density
is appropriate only in the case mentioned in item (iv) above.
The ontological interpretation is not, in essence, a probabilistic
interpretation.

d) The classical limit is very simple: we only have to find
the conditions for having $Q=0$.

e) There is no need to have a classical domain because this interpretation
is ontological. The question on why in a real measurement we do not see
superpositions of the pointer apparatus is answered by noting that,
in a measurement,
the wave function is a superposition of non-overlaping wave
functions \cite{dew2}. The particle will enter in 
one region, and it will be influenced by the unique quantum potential
obtained from the sole non-zero wave function defined on this region.

Of course this interpretation has still some flaws. It is difficult
to accomodate it with the notion of spin, it works only in the coordinate representation \cite{bola2},
its generalization to quantum fields is not yet completely understood
(see however \cite{kal}), just to mention some of them. 
Nevertheless, as it is an interpretation which does not 
require a classical domain, and which can be applied to a single system,
we think it should be relevant to examine what it can say about
quantum cosmology. 

\section{The application of the causal interpretation to quantum cosmology}

The hamiltonian of General Relativity (GR) without matter is given by:

\begin{equation}
\label{303}
H_{GR} = \int d^3x(N{\cal H}+N_j{\cal H}^j) ,
\end{equation}
where
\begin{eqnarray}
\label{39}
{\cal H} &=& G_{ijkl}\Pi ^{ij}\Pi ^{kl}-h^{1/2}R^{(3)} ,\\
\label{300}
{\cal H}^j &=& -2D_i\Pi ^{ij} .
\end{eqnarray}
The momentum ${\Pi}_{ij}$ canonically conjugated to 
the space metric $h^{ij}$ of the spacelike hypersurfaces which foliate
spacetime is

\begin{equation} 
\label{37}
\Pi _{ij} = \frac{\delta L}{\delta (\partial _t h^{ij})} = 
- h^{1/2}(K_{ij}-h_{ij}K) ,
\end{equation}
where
\begin{equation} 
K_{ij} = -\frac{1}{2N} (\partial _t h_{ij} -  \nabla _i N_j - \nabla _j N_i ) ,
\end{equation} 
and
\begin{equation}
\label{301}
G_{ijkl}=\frac{1}{2}h^{-1/2}(h_{ik}h_{jl}+h_{il}h_{jk}-h_{ij}h_{kl}),
\end{equation}
which is called the DeWitt metric. The quantity $R^{(3)}$ is the
intrinsic curvature of the hypersurfaces and $h$ is the determinant
of $h_{ij}$.
The lapse function $N$ and the shift function $N_j$ are the 
Lagrange multipliers of the super-hamiltonian constraint
${\cal H}$ and the super-momentum constraint ${\cal H}^j$,
respectively. They are present due to the invariance of GR under
spacetime coordinate transformations. Their specifications fix the 
coordinates.

If we follow the Dirac quantization procedure, these constraints
become conditions imposed on the possible states of the quantum
system, yielding the following quantum equations:

\begin{eqnarray}
\label{smo}
D_j\frac{\delta \Psi(h^{ij})}{\delta h^{ij}} &=& 0 \\
\label{wdw}
(G^{ijkl}\frac{\delta}{\delta h^{ij}}
\frac{\delta}{\delta h^{kl}}+h^{1/2}R^{(3)})\Psi(h^{ij}) &=& 0 
\end{eqnarray}
(we have set $\hbar =1$).

The first equation has a simple interpretation. It means that
the value of the wave function does not change
if the spacelike metric changes by a coordinate transformation.

The second one is the Wheeler-DeWitt equation, which should
determine the evolution of the wave function. However, time has
disappeared from it. There should
exist one momentum which is canonically conjugate to some intrinsic time
in which the quantum dynamics takes place. In
the time reparametrization invariant formulation of the quantum
mechanics of a non-relativistic particle, this particular momentum is easily
distinguishable from the others because it appears linearly 
in the quantum equation analogous to (\ref{wdw}), while the 
others appear quadratically.
However, in equation (\ref{wdw}), there is no momentum which
appears linearly; all of them appear quadratically. Hence, where
is time? This is the famous issue of time. This fact makes people advocates another quantization scheme, the ADM
approach, where time is chosen before quantization by a gauge 
fixing procedure. However,
different choices of time lead to inequivalent quantum theories 
\cite{isham,lemos} 
and there is no criterium to choose one of them.

Others say that the fact that it is not easy to find what should play
the role of time in the Wheeler-DeWitt equation simply means
that there is no time at all in quantum gravity \cite{barb1,barb2}. In fact, 
the good analogy with the time
reparametrization invariant quantum mechanics of non-relativistic
particles is via the Jacobi action:

\begin{equation}
\label{jac}
S = \int d\tau \sqrt{F_{E} T},
\end{equation}
where $F_{E} \equiv E - V$ and $T = \frac{1}{2} \sum _{i=1} ^{n}
m_{i} \frac{dx^{i}}{d\tau}\frac{dx^{i}}{d\tau}$. This is the
appropriate action when a closed conservative system is studied.
The conserved energy is $E$, and $V$ and $T$ are the potential
and kinetic energies of the system. This action yields 
Newton's equations of motion if a suitable choice of the parameter
$\tau$ is made such that $T = F_{E}$. 
The hamiltonian can be
calculated in the same way as before and it turns out to be
proportional to the following constraint:  
\begin{equation}
\label{tind}
\frac{1}{2} \sum _{i=1} ^{n} \frac{p^i p^i}{m_i} - F_{E} \approx 0 .
\end{equation}
Following the Dirac quantization scheme, this constraint yields
the following quantum equation:
\begin{equation}
\label{tind2}
\frac{1}{2} (\sum _{i=1} ^{n} \frac{\hat{p}^i \hat{p}^i}{m_i} +
V) \Psi (x^i) = E \Psi (x^i),
\end{equation}
which is the time independent Schr\"{o}dinger equation.
This is the correct analogous equation to the Wheeler-DeWitt 
equation (\ref{wdw}) because it is also quadratic in all
momenta. Consequently, we should consider the Wheeler-DeWitt
equation as a time-independent Schr\"{o}dinger equation with
zero energy. This is consistent with the fact that a closed
Universe has, by definition, a null total energy. 

Using a non-ontological interpretation, we can understand this
fact in another way. Space geometry is like position in ordinary
particle mechanics while spacetime geometry is like a trajectory.
Trajectories have no physical meaning in the quantum mechanics of
particles following a non-epistemological interpretation.
Instantaneous positions have. Analogously, 
spacetime has no physical meaning in quantum gravity, only space geometries
have. Hence, time makes no sense at the Planck scale. Space is the 
most primitive concept \cite{barb1,barb2}.
Therefore, it is quite natural that the Wheeler-DeWitt equation
of closed spaces be time independent. It is a time independent
Schr\"{o}dinger equation for zero energy, as it should be!

However, if we apply the ontological interpretation to quantum
cosmology, we should expect that the notion of a spacetime
would have a meaning exactly like the notion of trajectories
have in the causal interpretation of quantum mechanics of 
non-relativistic particles.
Hence, we should expect that the notion of time would emerge
naturally in this interpretation. Indeed, following the steps
we made in order to describe the ontological interpretation
in the beginning of this section, we substitute $\Psi = R \exp (iS/\hbar)$
into the Wheeler-DeWitt equation (\ref{wdw}), yielding the two equations
(for simplicity we stay in pure gravity):
\begin{equation}
\label{oqg}
\frac{1}{2}G_{ijkl}\frac{\delta S}{\delta h_{ij}}
\frac{\delta S}{\delta h_{kl}}-h^{1/2}R^{(3)}(h_{ij}) +
h^{1/2}Q(h_{ij}) = 0 ,
\end{equation}
\begin{equation}
\label{opr}
G_{ijkl}\frac{\delta}{\delta h_{ij}}(R^2
\frac{\delta S}{\delta h_{kl}}) = 0 ,
\end{equation}
where the quantum potential is given by:
\begin{equation}
\label{qgqp}
Q = -\frac{1}{R} G_{ijkl}\frac{\delta ^2 R}
{\delta h_{ij} \delta h_{kl}} .
\end{equation}

As before, we postulate that $h^{ij}(x,t)$ is meaningful even
at the Planck length and set: 
\begin{equation} 
\label{mqg}
\Pi _{ij} = 
- h^{1/2}(K_{ij}-h_{ij}K) = \frac{\delta S}{\delta h^{ij}} ,
\end{equation}
recalling that
\begin{equation}
\label{extr} 
K_{ij} = -\frac{1}{2N} (\partial _t h_{ij} -  \nabla _i N_j - \nabla _j N_i ) . 
\end{equation}
Hence, as $K_{ij}$ is essentially the time derivative of
$h_{ij}$, equation (\ref{mqg}) gives the time evolution
of $h_{ij}$. This time evolution will be different from the classical one
due to the presence of the quantum
potential in equation (\ref{oqg}), which may prevent, among other things,
the formation of classical singularities. 
The notion of spacetime is meaningful in this interpretation,
exactly like the notion of trajectory is meaningful in particle
quantum mechanics following this interpretation. However, it is not clear if
the spacetime geometries constructed from the non-classical solutions
$h_{ij}(x,t)$ of equations (\ref{oqg}-\ref{extr}) with different choices
of $N(x,t)$ and $N_i(x,t)$ will be the same, as in the classical case.
This problem will be discussed in more details in the last section. 

In the case of homogeneous models, however, the
supermomentum constraint ${\cal H}^i$ is identically zero, and the shift
function $N_i$ can be set to zero in equation (\ref{303}) without loosing
any of the Einstein's equations. The hamiltonian (\ref{303}) is
reduced to:
\begin{equation} 
\label{homham}
H_{GR} = N(t) {\cal H}(p^{\alpha}(t), q_{\alpha}(t)),
\end{equation}
where $p^{\alpha}(t)$ and $q_{\alpha}(t)$ represent the homogeneous 
degrees of freedom coming from $\Pi ^{ij}(x,t)$ and $h_{ij}(x,t)$.
Equations (\ref{oqg}-\ref{extr}) become:
\begin{equation}
\label{hoqg}
\frac{1}{2}f_{\alpha\beta}(q_{\mu})\frac{\partial S}{\partial q_{\alpha}}
\frac{\partial S}{\partial q_{\beta}}+ U(q_{\mu}) +
Q(q_{\mu}) = 0,
\end{equation}
\begin{equation}
\label{hqgqp}
Q(q_{\mu}) = -\frac{1}{R} f_{\alpha\beta}\frac{\partial ^2 R}
{\partial q_{\alpha} \partial q_{\beta}},
\end{equation}
\begin{equation}
\label{h}
p^{\alpha} = \frac{\partial S}{\partial q_{\alpha}} =
f^{\alpha\beta}\frac{1}{N}\frac{\partial q_{\beta}}{\partial t} = 0,
\end{equation}
where $f_{\alpha\beta}(q_{\mu})$ and $U(q_{\mu})$ are the minisuperspace
particularizations of $G_{ijkl}$ and $-h^{1/2}R^{(3)}(h_{ij})$, respectively.
 
Equation (\ref{h}) is invariant under time reparametrization. Hence,
even at the quantum level, different choices of $N(t)$ yield the same 
spacetime geometry for a given non-classical solution $q_{\alpha}(x,t)$.

\section{The singularity problem}

The question about the persistency of classical cosmological
singularities at the quantum level for homogeneous fields 
has been studied extensively
in the literature. In a first approach, the dynamical evolution
of the quantum states is obtained by fixing the time gauge before
quantization. As we mentioned above, different
choices of time gauge imply different quantum theories with
different answers to the question we are addressing \cite{isham,lemos}.
In the last section we have shown that this ambiguity in the choice
of time does not arise if we apply the causal interpretation to
quantum cosmology in the case of minisuperspace models of
homogeneous fields. In the present section, we will bring home this 
fact by making use of a simple minisuperspace example, where
the existence of cosmological
singularities at the quantum level does not depend on the choice
of the time-gauge but only on the choice of the quantum state of the
system. This minisuperspace is the Bianchi I model.

The minisuperspace metric is given by:

\begin{eqnarray}
\label{bia}
ds^2 &=& -N^2(t)dt^2+ 
\exp [2\beta_0(t) + 2\beta_+(t) + 2\sqrt{3}\beta_-(t)] \; dx^2 + \nonumber \\
& & \exp [2\beta_0(t) + 2\beta_+(t) - 2\sqrt{3}\beta_-(t)] \; dy^2 +
\nonumber \\
& & \exp [2\beta_0(t) - 4\beta_+(t)] \; dz^2
\end{eqnarray}
The gravitational hamiltonian for this minisuperspace model is:

\begin{equation}
\label{hbia}
{\cal H} = \frac{N}{24 \exp{(3\beta _0)}}(p_0^2 - p_+^2 - p_-^2).
\end{equation}
where the $p$'s are the canonical momenta of the $\beta$'s.
The classical equations of motion are:

\begin{equation}
\label{hbia1}
p_0^2 - p_+^2 - p_-^2 = 0 ,
\end{equation}

\begin{equation}
\label{hbia2}
\dot{\beta _0} = \frac{\partial{\cal H}}{\partial p_0} = 
\frac{N}{12 \exp{(3\beta _0)}} p_0 ,
\end{equation}

\begin{equation}
\label{hbia3}
\dot{\beta _+} = \frac{\partial{\cal H}}{\partial p_+} = 
-\frac{N}{12 \exp{(3\beta _0)}} p_+ ,
\end{equation}

\begin{equation}
\label{hbia4}
\dot{\beta _-} = \frac{\partial{\cal H}}{\partial p_-} = 
-\frac{N}{12 \exp{(3\beta _0)}} p_- ,
\end{equation}

\begin{equation}
\label{hbia5}
\dot{p_0} = -\frac{\partial{\cal H}}{\partial \beta _0} = 
-\frac{N}{8 \exp{(3\beta _0)}}(p_0^2 - p_+^2 - p_-^2) = 0,
\end{equation}

\begin{equation}
\label{hbia6}
\dot{p_+} = -\frac{\partial{\cal H}}{\partial \beta _+} = 0 ,
\end{equation}

\begin{equation}
\label{hbia7}
\dot{p_-} = -\frac{\partial{\cal H}}{\partial \beta _-} = 0 .
\end{equation}

To discuss the appearance of singularities, we need the Weyl square
tensor $W^2 \equiv W^{\alpha\beta\mu\nu} W_{\alpha\beta\mu\nu}$. 
It is given by:
\begin{equation}
\label{w2}
W^2 = \frac{1}{432} e^{-12 \beta _0} (-2p_0 p_+^3 + 6 p_0 p_-^2 p_+ +
p_+^4 + 2 p_+^2 p_-^2 + p_+^4 + p_0^2 p_+^2 + p_0^2 p_-^2).
\end{equation}
Hence, the Weyl square tensor is proportional to
$\exp{(-12\beta _0)}$ because the $p$'s are constants 
(see Eqs (\ref{hbia5}-\ref{hbia7})). Solving equation (\ref{hbia2}) in the gauge 
$N=12\exp (3 \beta _0)$, we can see that
$\beta _0=p_0 t$, and the singularity is at $t=-\infty$. It is 
a fast-time gauge in the terminology of reference \cite{lemos}. 
If we choose $N=1$, then $\beta _0=\frac{1}{3} 
\ln({\frac{p_0}{4} t})$ and the singularity appears at $t=0$. It is 
a slow-time gauge.
The classical singularity can be avoided only if we set $p_0 =0$.
But then, due to equation (\ref{hbia1}), we would also have $p_{\pm} =0$,
implying that the Weyl square tensor be identically zero,
corresponding to the trivial case of Minkowski spacetime.
The conjecture stated in reference \cite{lemos} says that the singularity 
persists at the quantum level in the fast-time gauge but disappears in the
slow-time gauge.

The Dirac quantization scheme yields the following Wheeler-DeWitt equation:

\begin{equation}
\label{wbia}
\left(\frac{\partial ^2} {\partial \beta_0 ^2} -
\frac{\partial ^2} {\partial \beta _+ ^2} -
\frac{\partial ^2} {\partial \beta _- ^2}\right)\Psi =0.
\end{equation}
In reference \cite{ash}, a consistent inner product is constructed,
and gauge invariant (Dirac) observables which dependes on a parameter,
which is nothing but $\beta _0$, are constructed. In this way,
the Weyl square observable is built, exhibiting a singularity
at $\beta _0=-\infty$, as in the classical case. As $\beta _0$
plays the role of time, this is equivalent to a quantization
in the fast-time gauge.

Let us now make use of the causal interpretation.
Take the following solution to the
Wheeler-DeWitt equation (\ref{wbia}):

\begin{eqnarray}
\label{obia}
\Psi &=& \exp{[i(\sqrt{k_+^2 + k_-^2} \; \beta _o + k_+ \, \beta _+ +
 k_- \, \beta _-)]} + \nonumber \\
& &  \exp{[-i(\sqrt{l_+^2 + l_-^2} \; \beta _o + l_+ \, \beta _+ + 
l_- \, \beta _-)]}
\end{eqnarray}
where the $k$'s and $l$'s are real constants.
Note that this function is not normalizable, but this is not important
for the ontological interpretation.
Calculating $\frac{\partial S}{\partial \beta _0}$, 
$\frac{\partial S}{\partial \beta _+}$, and
$\frac{\partial S}{\partial \beta _-}$, where $S$ is the phase
of the wave function (\ref{obia}), we obtain:

\begin{equation}
\label{obia1}
p_0 \equiv  \frac{\partial S}{\partial \beta _0} =
\frac{1}{2}\sqrt{k_+^2 + k_-^2} - \frac{1}{2}\sqrt{l_+^2 + l_-^2},
\end{equation}

\begin{equation}
\label{obia2}
p_+ \equiv \frac{\partial S}{\partial \beta _+} = 
\frac{1}{2}k_+ - \frac{1}{2}l_+ ,
\end{equation}

\begin{equation}
\label{obia3}
p_- \equiv \frac{\partial S}{\partial \beta _-} = 
\frac{1}{2}k_- - \frac{1}{2}l_- .
\end{equation}

It is easy to see in the above equations that is possible to
have $p_0 = 0$ and $p_{\pm} \neq 0$. We can also understand it
by the fact that equation (\ref{hbia1}) is no longer valid at the
quantum level; the quantum potential must be added to it and thus
$p_0 = 0$ does not imply $p_{\pm} = 0$. Hence, it is possible to find
a curved spacetime without singularities, i.e., a spacetime with a 
Weyl square 
tensor which is neither null nor infinite, for the quantized Bianchi I 
model. Note that this result is independent on the value chosen
for $N$. In particular, we could have chosen the fast-time gauge mentioned
previously, and still have a non-singular quantum spacetime.
Hence, using the ontological interpretation, we have presented a
simple example where the appearance of singularities in the quantum
regime depends only on the state of the system, and not on the
time-gauge choice we make.

\section{Conclusion}

In this paper we have shown that the application of the causal 
interpretation of quantum mechanics to the quantum cosmology of
homogeneous fields yields definite predictions without any ambiguity
due to the arbitrariness in the time-gauge choice. As a consequence,
the slow-fast-time gauge conjecture about the persistency of cosmological
singularities at the quantum level is irrelevant within the causal
interpretation. Taking the minisuperspace
of Bianchi I model, we have shown
that the quantum potential of given quantum states can prevent the
formation of the classical singularity, yielding a non-trivial regular 
four-geometry, independently on the choice of the lapse function.

One can argue on why we have obtained non-singular quantum solutions
in the quantization of the Bianchi I model while in reference \cite{ash}
it is shown that all quantum states of this model are singular. The
answer is that in order to have a Dirac observable which can plays
the role of time, Ashtekar {\it et al}. \cite{ash} had to take only
positive frequency solutions of equation (\ref{wbia}), i.e., states with 
positive $p_0$. In this way, the Dirac observable ${\hat{\beta}}_0$
becomes proportional to the identity operator, the multiplying constant
being time. In the causal interpretation, however, the restriction
to positive $p_0$ is not necessary for having a notion of time: it
appears, as in the classical case, via equation (\ref{h}). Hence, we can 
construct wave functions which are
superpositions of positive and negative frequency solutions as in 
equation (\ref{obia}), and which does not present any singularity, as was
demonstrated in the last section. Note that any superposition of
positive and negative solutions is not an eigenstate of the operator 
${\hat{\beta}}_0$ of reference \cite{ash}. Indeed, if the Hilbert space 
is enlarged with the inclusion of negative frequency solutions, 
we cannot use ${\hat{\beta}}_0$
as a time operator because it is no longer a multiple of the identity.

A very interesting and fundamental question is about the generalization 
of this result to the general case of inhomogeneous fields. 
In this case, the
supermomentum constraint ${\cal H}^i$ is not identically zero, and the shift
function $N_i$ must be present in the hamiltonian (\ref{303}). The simple
time-reparametrization invariant equation (\ref{h}) will no longer be valid.
We have to use the general equations (\ref{mqg}) and (\ref{extr}), where 
$S$ is a solution of the modified Hamilton-Jacobi equation (\ref{oqg}). 

One can see the problem more clearly by trying to construct a 
hamiltonian which generates the non-classical 
evolution of $h_{ij}$. It would
be given by the hamiltonian (\ref{303}), with ${\cal H}$ suplemented by
the quantum potential (\ref{qgqp}): 
\begin{equation}
\label{n303}
{\tilde{H}}_{GR} = \int d^3x(N \tilde{{\cal H}}+N_j{\cal H}^j) 
\end{equation}
where
\begin{equation}
\label{n39}
\tilde{{\cal H}} = {\cal H} - \frac{1}{R} G_{ijkl}\frac{\delta ^2 R}
{\delta h_{ij} \delta h_{kl}} 
\end{equation}

However, it is not clear if the Poisson brackets of the constraints
$\tilde{{\cal H}}$ and ${\cal H}^j$ close, and it is sure that they do
not close like the commutators of the generators of the deformations of
three-dimensional spacelike slices cut through a Riemannian spacetime.
Indeed, in reference \cite{hoj} it is shown that the potential term in the
super-hamiltonian must be proportional to the scalar curvature of the 
spacelike slices plus a cosmological term, exactly like in General Relativity,
in order that the dynamics of the fields be consistent with the kinematic 
of deformations. Hence, the dynamics of $h_{ij}$ in
the presence of the quantum potential does not satisfy this requirement.
This is a very important point which should be investigated further.

Note that, for homogeneous quantum cosmology, the non classical evolution
of the homogeneous degrees of freedom can be generated by a hamiltonian
with a single constraint,
\begin{equation} 
\label{nhomham}
{\tilde{H}}_{GR} = N(t) \tilde{{\cal H}}(p^{\alpha}(t), q_{\alpha}(t)),
\end{equation}
where $\tilde{{\cal H}}(p^{\alpha}(t), q_{\alpha}(t))$ is the classical 
constraint suplemented by the quantum potential
\begin{equation}
\label{hqgqp2}
Q(q_{\mu}) = -\frac{1}{R} f_{\alpha\beta}\frac{\partial ^2 R}
{\partial q_{\alpha} \partial q_{\beta}}.
\end{equation}
As a single constraint commutes with itself, the theory is invariant under time
reparametrizations and the problems mentioned above do not arise in this
restricted case.

\section*{Acknowledgments}
Part of this work was done while NPN was a PREVI (Special Program for
Visiting Professor/Researcher) fellow at the Physics Department of the
Federal University at Juiz de Fora.
We would like to thank the
group of the ``Pequeno Semin\'ario" at CBPF for useful discussions and FAPEMIG
for financial support. NPN
would like to thank CNPq for financial support and the Federal
University at Juiz de Fora, UFJF, for hospitality. JAB would like to thank
the Laboratory for Cosmology and Experimental High Energy Physics 
(Lafex) at the
Brazilian Center for Physical Research (CBPF/CNPq) for hospitality.

\end{document}